\documentclass[aps, prd, amsmath, floats, floatfix, twocolumn, nofootinbib, showpacs]{revtex4-1}
\usepackage{graphicx}
\usepackage{amssymb}
\usepackage{mathrsfs}
\usepackage{color}

\newcommand{\be}{\begin{eqnarray}}
\newcommand{\ee}{\end{eqnarray}}

\begin{document}

\title{A note on the observational evidence for the existence of event horizons\\ in astrophysical black hole candidates}

\author{Cosimo Bambi}
\email{bambi@fudan.edu.cn}
\affiliation{Arnold Sommerfeld Center for Theoretical Physics,
Ludwig-Maximilians-Universit\"at M\"unchen, D-80333 Munich, Germany}
\altaffiliation{Present address: Department of Physics,
Fudan University, Shanghai 200433, China}

\date{\today}

\begin{abstract}
Black holes have the peculiar and intriguing property of having an event
horizon, a one-way membrane causally separating their internal region from
the rest of the Universe. Today astrophysical observations provide some 
evidence for the existence of event horizons in astrophysical black hole 
candidates. In this short paper, I compare the constraint we can infer from the 
non-observation of electromagnetic radiation from the putative surface of 
these objects with the bound coming from the ergoregion instability,
pointing out the respective assumptions and limitations.
\end{abstract}


\maketitle

%%%%%%%%%%%%%%%%%%%%%%%%%%%%%%%

\section{Introduction}

A black hole (BH) can be defined as the region $\mathcal{B}$ of the total 
space-time $\mathcal{M}$ which does not overlap with the causal past 
of future null infinity $J^- (\mathscr{I}^+)$~\cite{poi}: 
\be
\mathcal{B} = \mathcal{M} - J^- (\mathscr{I}^+) \, . 
\ee
The {\it event horizon} of a BH is the boundary delimiting the BH. 
Everything falling onto the BH and crossing the event horizon is lost 
for ever and it cannot affect events happening outside the BH any more. 
However, it may be possible that event horizons never form in nature, but 
that only apparent horizons can be created~\cite{h}. An {\it apparent horizon} 
is a closed surface of zero expansion for a congruence of outgoing null 
geodesics orthogonal to the surface~\cite{poi}. Outward-pointing light rays 
behind an apparent horizon actually move inwards and therefore they cannot cross the 
apparent horizon. In the special case of a stationary space-time, an event 
horizon is also an apparent horizon, but the reverse is not true in general. 
In particular, the event horizon is determined by the global properties of 
the space-time, while the apparent horizon depends on the observer.

Astronomers have discovered at least two classes of astrophysical BH
candidates~\cite{narayan}: stellar-mass objects in X-ray binary systems and 
super-massive objects at the center of every normal galaxy. These objects 
are thought to be BHs because they cannot be explained otherwise without 
introducing new physics: the stellar-mass BH candidates are too heavy to be 
neutron stars for any reasonable matter equation of state~\cite{ns}, while at 
least some of the super-massive objects in galactic nuclei are too heavy, 
compact, and old to be clusters of non-luminous bodies~\cite{maoz}. 
There is also a set of observations suggesting that BH candidates have really
an event horizon~\cite{quiescent,outburst,sgra}. Basically, these objects 
seem to be able to swallow all the accreting gas without emitting any kind of 
electromagnetic radiation from their putative surface. In the case of low-mass
X-ray binaries, we can compare systems in which the primary is thought 
to be a BH and the ones in which the primary is thought to be a neutron star.
In the quiescent state, we can observe thermal radiation from the 
surface of neutron stars, while no such a radiation is observed from BH 
candidates~\cite{quiescent}. Neutron star systems show type-I X-ray bursts
(as outcome of compression and heating of the gas accumulated on their 
surface), while the phenomenon has never been observed in binaries with BH 
candidates~\cite{outburst}. There are also strong constraints on the radiation 
emitted by the possible surface of the supermassive BH candidate at the 
center of our Galaxy~\cite{sgra}.

This body of observations can be easily explained with the fact that BHs
have no surface and that the gas crossing the event horizon cannot
be seen by distant observers any more (see however Ref.~\cite{abra}). 
Strictly speaking, the confirmation 
for the existence of an event horizon would require the knowledge of the 
future null infinity of the Universe, which is clearly impossible for us. On the
contrary, the non-observation of electromagnetic radiation emitted by the 
gas after falling into the compact object nicely meets the definition of 
apparent horizon. However, the geometry of the space-time around 
astrophysical BH candidates is practically stationary for the timescale 
of our observations, and that may make impossible to discriminate an 
event horizon from an apparent horizon.

\section{Electromagnetic constraint}

Let us image a BH as a gas of particles packed in a small region by 
the gravitational force\footnote{The model of BH I will consider may remind
the one discussed in Ref.~\cite{gia}. The radius of the compact object, 
$R$, is larger than the one corresponding to the event horizon of a 
(classical) BH with the same mass and spin.}. 
As this gas has a finite temperature, it must radiate. 
However, if the object is very compact, the emitted radiation is strongly 
redshifted when it reaches a distant observer and the object can appear 
very faint. Here, I relax the quite common assumption of steady state 
$L = \dot{M} c^2$~\cite{quiescent,sgra}, where $L$ is the surface luminosity 
and $\dot{M}$ is the mass accretion rate. That would requires that the 
accreting gas hits the ``solid surface'' of the object and then radiates to infinity 
all its kinetic energy. If this were the case, a very compact object would not be 
able to increase its mass, or at least the process would be very inefficient, 
likely in contradiction with the observations of the super-massive objects 
in galactic nuclei. Moreover, there are no reasons to assume that BH 
candidates have a solid surface. In the picture in which we have a gas of 
particles packed in a small region by the gravitational force, the accreting 
gas enters into the compact object and both its rest-mass and kinetic 
energy contribute to increase the mass of the BH candidate.

Let us now see the constraint we can obtain in this picture from the 
non-observation of thermal spectrum from BH candidates. The specific 
energy flux density of the compact object (often measured in 
erg~cm$^{-2}$~s$^{-1}$~Hz$^{-1}$) as detected by a distant observer is 
\be
F = \int I_{\rm o} d\Omega \, ,
\ee
where $I_{\rm o}$ is the specific intensity of the radiation as measured by the
distant observer and $d\Omega$ is the element of the solid angle subtended
by the image of the object on the observer's sky. $I_{\rm x}/\nu_{\rm x}^3 = 
{\rm const.}$ (Liouville's Theorem), where $\nu_{\rm x}$ is the photon frequency 
measured by any local observer on the photon path, and
\be
d\Omega = \frac{dxdy}{D^2} \, ,
\ee
where $x$ and $y$ are the Cartesian coordinates on the observer's sky
and $D$ is the distance of the compact object from the observer. The 
{\it equivalent isotropic luminosity} of the BH candidate is thus
\be
L = 4 \pi \int g^3 I_{\rm e} dx dy d\nu \, .
\ee
Here $g = \nu_{\rm o}/\nu_{\rm e}$ is the redshift factor, $\nu_{\rm o}$ is the 
photon frequency measured by the distant observer, and $\nu_{\rm e}$ and
$I_{\rm e}$ are respective the photon frequency and the specific intensity of 
the radiation measured by an observer located at the point of emission of
the photon, on the surface of the compact object, and corotating with the 
surface of the compact object. The emission should be like the one of a
blackbody; that is,
\be
I_{\rm e} = \frac{2h\nu^3_{\rm e}}{c^2}\frac{1}{\exp
\left(\frac{h\nu_{\rm e}}{k_{\rm B}T_{\rm e}}\right) - 1} \, ,
\ee
where $T_{\rm e}$ is the temperature of the surface of the BH candidate 
measured by a locally corotating observer.

For the sake of simplicity, we now consider a spherically-symmetric non-rotating 
object. The geometry of the space-time around the BH candidate will be described by
the Schwarzschild solution, which is valid till the radius of the compact 
object, $R$. The luminosity becomes
\be\label{eq-l}
L = 4 \sigma g^4 T_{\rm e}^4 \int dx dy \, ,
\ee
where $\sigma$ is the Stefan-Boltzmann constant and
\be
g = \left( 1 - \frac{2M}{R} \right)^{1/2} \, .
\ee
Here $g$ is a constant, but it would be a function of $x$ and $y$ in a 
more general background. The integrand in Eq.~(\ref{eq-l}) is simply the area of the 
apparent image of the BH candidate on the observer's sky:
\be
\int dxdy = \pi R_{\rm app}^2 = 
\left\{
\begin{array}{ll}
27 \pi M^2 & R < 3 M \\
\pi \frac{R^2}{g^2} & R > 3 M
\end{array} 
\right. \, .
\ee
The radius $r = 3M$ is the capture photon radius of the Schwarzschild
space-time. Inside such a radius, the gravitational force is so strong that
any light rays coming from infinity is captured by the compact object.

A distant observer sees therefore an object with an apparent temperature
\be
T_{\rm app} = g T_{\rm e} \approx T_{\rm e} 
\left(\frac{\delta}{2M}\right)^{1/2} \, ,
\ee
where I wrote $R = 2M + \delta$ and assumed $\delta \ll 1$ and positive. The most stringent 
constraint on $\delta$ can be inferred from the observations of the supermassive 
BH candidate at the center of our Galaxy. Infrared and near-infrared data
require $T_{\rm app} < 0.01$~eV~\cite{sgra}. If we assume a local temperature as 
high as $k_{\rm B} T_{\rm e} \sim m_{\rm p} c^2 \sim 1$~GeV (roughly the
gravitational binding energy of a proton), we find
\be\label{eq-c-em}
\delta < 10^{-10} \; {\rm cm} \, ,
\ee
as $M \approx 6 \cdot 10^{11}$~cm. With a lower temperature $T_{\rm e}$,
the constraint would be weaker, while a higher temperature 
seems to be unlikely, as the object is old and the accreting gas would have
already cooled it down. The proper distance of the boundary of the
BH candidate from the event horizon of a Schwarzschild BH with the same 
mass is 
\be\label{eq-c-em2}
\Delta \approx \frac{\delta}{\sqrt{1 - \frac{2M}{R}}} \approx 
\sqrt{2 M \delta} < 10 \; {\rm cm} \, .
\ee
Such a result should not change significantly if we consider a rotating object.

\section{Stability constraint}

The existence of event or apparent horizons in astrophysical BH candidates is
also suggested by considerations concerning the stability of these objects. It
is well known that rapidly-rotating very-compact objects may be affected by the
{\it ergoregion instability}~\cite{ergo}. In the ergoregion, $g_{tt} > 0$ (if the 
metric has signature $-+++$) and the frame-dragging is so strong that stationary
orbits are not allowed. That implies that in the ergoregion there are excitations
with negative energy with respect to a stationary observer at infinity. These
excitations can be seen as quasi-bound states: they are trapped by the gravitational
potential on one side, and by the surface of the object (or by the center of the
object if the latter is made of matter non-interacting with the excitations) on the 
other side. As some modes can escape to infinity carrying positive energy, negative
energy modes in the ergoregion can grow indefinitely, thus generating an 
instability. Objects with a horizon may instead be stable because there may not 
be quasi-bound states in the ergoregion: any excitation in the ergoregion is
swallowed by the BH. Let us notice, however, that the existence of a horizon
is not sufficient in general to prevent the ergoregion instability~\cite{pani}.

Roughly speaking, the instability timescale $\tau$ decreases as the angular 
velocity and the compactness of the compact object increases. For 
rotating very-compact objects, one typically finds that the instability
is strong and occurs on a dynamical timescale $\tau \sim M$~\cite{cardoso};
that is, $\sim 1$~s for objects with a mass $M \sim 10$~$M_\odot$ and
$\sim 10^7$~s if $M \sim 10^8$~$M_\odot$. While there are counter-examples
in which rotating compact objects can be stable or very-long living~\cite{rezzolla},
it seems difficult that the latter can still meet observations requiring that 
astrophysical BH candidates can rotate very rapidly~\cite{spin,spin2}
and have a high radiative efficiency~\cite{eta}.
Let us notice, however, that the issue of the ergoregion instability can be
discussed only within a well defined theoretical model (gravity theory, internal 
structure and composition of the compact object, etc.) and that it has been
studied only for a very limited number of specific cases. Considerations on the
non-observations of electromagnetic radiation from the surface of BH candidates
are much more model-independent and rely on a set of assumptions that can
be violated only invoking very exotic new physics.

Here, I will discuss the ergoregion instability within the following picture.
I assume that the geometry around an astrophysical BH candidate is
exactly described by the Kerr solution up the radius of the compact object, 
$R$. Considerations on the ergoregion instability indeed require a
specific background and we may think that possible deviations from the 
Kerr metric can be tested with other approaches~\cite{review}. In the case of a reflecting
surface, the timescale for scalar instabilities can be estimated as~\cite{shinji}
\be
\tau \sim A(M,a_*) \left| \ln \left( 
\frac{R - R_{\rm H}}{2M \sqrt{1 - a^2_*}} \right) \right| \, ,
\ee
where $R_{\rm H} = M(1 + \sqrt{1 - a_*^2})$ is the radius of the event horizon
of a Kerr BH with mass $M$ and spin parameter $a_*$. $A(M,a_*)$ is a 
function of $M$ and $a_*$. For moderate values of the spin 
parameter $a_*$, $A \sim M$; that is, the instability occurs on a dynamical
timescale. For high values of $a_*$, $A$ decreases very quickly. In the
case of a Kerr BH, $R = R_{\rm H}$ and the object is stable. On the other
hand, if $R = R_{\rm H} + \delta$, the fact that we observe
long-living rapidly rotating BH candidates demands
\be\label{eq-ei}
\delta, \; \Delta \ll L_{\rm Pl} \approx 10^{-33} \, {\rm cm} \, ,
\ee
where $\Delta$ is the physical distance encountered in the previous section.
Eq.~(\ref{eq-ei}) essentially rules out the possibility that current BH candidates have
no horizon, or at least something that behaves very much like a horizon 
for the unstable modes. The possibility of an exact Kerr background with $\delta$
so large that there is no ergoregion seems to be unlikely, as we know
objects that, when the space-time around them is described by the Kerr
solution, would have an accretion disk with inner edge inside the
ergosphere~\cite{spin}.

\section{Conclusions}

In conclusion, we have observations suggesting that BH candidates have
a horizon or at least putting constraints on the possible distance between the
boundary of these compact objects and the event horizon of a BH with the
same mass and spin. Such a distance can be seen as a measurement of 
how much close the formation of the horizon is. From the non-observation of 
thermal radiation from the putative surface of astrophysical BH candidates, 
one can infer the constraint in Eqs.~(\ref{eq-c-em}) and 
(\ref{eq-c-em2}): actually, such a bound is not so stringent, as one may argue 
that new physics can show up at much shorter scales. However, the 
result seems to be quite robust -- it is just supposed that the 
compact object must emit electromagnetic radiation due to its finite 
temperature -- and very exotic new physics is necessary to change these 
conclusions or to get a different bound. Considerations on the ergoregion
instability are instead to be taken with caution. The timescale instability 
strongly depends on the exact model; i.e. gravity theory, internal structure
and composition of the object, and so on, which we do not know. However, we
can optimistically arrive at the following conclusion. If the geometry around
astrophysical BH candidates is very close to the Kerr solution, the existence
of stable or long-living objects likely requires some kind of horizon.
Otherwise, we can probably hope to discover deviations from the Kerr
background with tests already proposed in the literature and possible in a near
future with new observational facilities.

%%%%%%%%%%%%%%%%%%%%%%%%%%%%%%%

\begin{acknowledgments}
This work was supported by the Humboldt Foundation.
\end{acknowledgments}

%%%%%%%%%%%%%%%%%%%%%%%%%%%%%%

\end{document}